\documentclass[pre,twocolumn,aps,showpacs]{revtex4} 
\usepackage{dcolumn}
\usepackage{graphicx}
\usepackage{graphicx,amssymb,amsmath}
\usepackage{natbib}
\usepackage{bm}

\begin{document}

\title{Effect of external stress on the thermal melting of DNA}

\author{Joseph Rudnick and Tatiana Kuriabova}

\affiliation{Department of Physics, UCLA, Box 951547, Los Angeles,
CA 90095-1547}

\date{\today}

\begin{abstract}
We discuss the effects of external stress on the thermal denaturation of homogeneous DNA.  
Pulling double-stranded DNA at each end exerts a profound effect on the thermal denaturation, or melting, of a long segment of this molecule. We discuss the effects on this transition of a stretching force applied to opposite ends of one of the DNA strands, including full consideration
of the consequences of excluded volume, the analysis of which  is greatly simplified in this case.
We also discuss the interplay of thermal denaturation and force-generated separation when the tension is generated by a force at the end of the duplexed strands and an equal and opposite force is applied to the other end of the second strand. 
\end{abstract}

%\pacs{not yet}

\maketitle

\section{Introduction} \label{sec:intro}

When heated, double-stranded DNA develops regions of strand separation, known as ``denaturation bubbles.'' \cite{psbook} As the temperature increases, these bubbles grow in size and proliferate. The culmination of this process is complete separation of the strands, that is the complete thermal denaturation---or melting---of DNA, a transition of significant biological and technological significance. Among the most important theoretical approaches to this process is a collection based, or mathematically related, to an underlying model introduced by Poland and Scheraga \cite{ps1,ps2}. These approaches include the formalism introduced by Peyrard and Bishop \cite{pb} that maps the process of melting into the disappearance of a bound state in a one-dimensional potential well. Given the results obtained with the use of this family of models, the melting process in the case of homogeneous double-stranded DNA---that is, the molecule consisting of, say, one strand  containing only cytosines and another containing only guanines---is well-understood. It is known with a relatively high degree of certainty that an infinitely long molecule of this kind will undergo a first order melting transition \cite{kmp}. This conclusion follows from the  consideration of the effects of self-avoidance---in particular, the consequences of self avoidance with regard to the structure of the vertex connecting an intact portion of DNA with a denaturation bubble. The treatment of this process follows from the work of Duplantier \cite{d1,d2} on the renormalization of vertices for arbitrary polymer networks. 

The sharp transition to the completely denatured state is an example of a phase transition in a one-dimensional system. The existence of such a transition follows from the effective long-range interaction inherent in the statistical mechanics of the denaturation bubbles. The correlations that are propagated in a bubble result in an effective Boltzmann factor consistent with the statistical mechanics of an inverse square Ising model, which is known to undergo a phase transition \cite{thouless1,yuvand}. 

%%Tatiana - I have commented out the sentence below because I can find no published work supporting it. Particularly, the work of Zocchi, Gruner et. al. has not been published.

%Recent experiments on homogeneous oligomers of DNA yield results that are consistent with this kind of behavior, in that they are consistent with the kind of incipient phase transition that one would expect in a finite system.

In light of the expectation of a sharp transition in thermodynamic limit in the case of homogeneous DNA, it is noteworthy that experiments on  biological DNA produce melting curves that are belie the expectation of a true phase transition in the thermodynamic limit. Here, we utilize the standard definition of a phase transition as non-analyticity in thermodynamic functions. What is seen experimentally is, rather a collection of highly-structured, but nevertheless smooth, curves when, for instance, the specific heat is measured (see, e.g. \cite{gotoh}). It appears that the inhomogeneity inherent in the DNA present in living organisms, which effectively translates to random inhomogeneity in the context of thermal denaturation, profoundly affects the nature of the transition. This is consistent with the fact that such random inhomogeneity is relevant in the sense of the Harris criterion \cite{harris,monthus}. A brief demonstration that this is so is presented in Appendix~\ref{sec:harris}.

An additional mechanism for the separation of DNA strands is the application of external forces, which under appropriate circumstances leads to the ``unzipping''  \cite{bensimon1} or to the stretching-induced ``melting'' of the molecule. This process has been investigated theoretically \cite{lubnels1,lubnels2,lubnels3,RandB1,RandB2} as has the interplay between the effects of externally applied force and thermal fluctuations. In the latter case, the melting process is modeled as a helix-coil transition, which can be represented in terms of a system with an intimate relationship to the one-dimensional Ising model with short range interactions. In this approach to the denaturation process, there is no prospect of a thermodynamic transition \cite{Landau:51}.
 
The set of calculations that we report here assumes the underlying validity of the Poland-Scheraga-based approaches to DNA melting. We consider melting as the result of the accumulation and possible merging of denaturation bubbles. What is added to the picture is a pair of equal and opposite forces at the two ends of the denaturing strands. Restricting our focus to homogeneous DNA, we assess the consequences of this force pair under two circumstances: \begin{enumerate}
\item[\bf{a}:] \label{item:first}The forces are applied either to both strands simultaneously or to both ends of one of them.
\item[\bf{b}:] \label{item:second} The force at one end acts on one strand, while the force at the opposite end acts on the other strand.
\end{enumerate}

\begin{figure}[htbp]
\begin{center}
\includegraphics[width=2.5in]{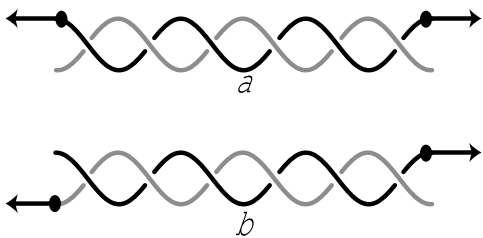}
\caption{The two ways in which a strand of duplexed DNA can be stretched, either {\bf{a}}: with forces applied to opposite ends of the same strand, or {\bf{b}}: with a force at the left end of one of the strands and an equal and opposite force applied to the right end of the other strand. }
\label{fig:stretch}
\end{center}
\end{figure}

These two ways of pulling at the ends of duplexed DNA are illustrated in Fig. 
\ref{fig:stretch}.

In the first case above, the effect of the force is two-fold. First, as noted by Rouzina and Bloomfield \cite{RandB1}, stretching both strands produces an energetic bias in favor of denaturation, in that single strands of DNA are more easily elongated than the intertwined strands of the duplexed version of the molecule. On the other hand, DNA melting represents the classic competition between energy and entropy, with denaturation bubbles embodying the entropically favored, energetically costly state. Stretching a pair of strands has the effect of applying an additional energy penalty to the states available to the separated strands. Thus, the imposition tension on melting strands of DNA can either \emph{promote} or \emph{inhibit} thermal denaturation. As we will see, both consequences can be observed.

The second version of an externally generated tension has the unequivocal effect of favoring strand separation. In fact, it is straightforward to demonstrate that pulling one strand at one end and the other strand at the opposite end causes the state of the molecule in which any portion is duplexed to be a metastable state, rather than a state of true equilibrium. 

The paper proceeds as follows. In Section~\ref{sec:extforce} we review the effect of stress on the statistics of a Gaussian chain of equal and opposite forces applied to both ends and introduce the mathematical tools for the analyzing the phase transition in terms of the nonanalyticity of the generating functions. In particular, we recall that the asymptotic statistics of a system with a fixed number of monomeric units is controlled in the thermodynamic limit by the singularity in the generating function lying closest to the origin in the complex fugacity plane \cite{elements}. Given this, we are able to show that almost all modifications of the generating function that follow from excluded volume considerations will exert a negligible effect on the melting transition, in that the singularities associated with those modifications are further away from the origin than the singularity that arises in the case of the unrestricted chain. The mitigation of these excluded volume effects can be simply understood in terms of the energy cost of a self intersection in light of the forces acting on the ends of the chain.   In Section~\ref{sec:bubble} we focus on the one way in which self-avoidance influences the statistical mechanics of the melting transition, in the interior of a denaturation bubble. We find that it introduces a logarithmic modification to the mean field result for melting exponents. Section~\ref{sec:effects} is devoted to a general analysis of the influence of self-avoidance and stress on thermal denaturation of DNA, particularly as it relates to the phase diagram and key temperature dependences. In Section~\ref{sec:pulling} we present a preliminary analysis of the consequences of stress applied as in Fig. \ref{fig:stretch}b. We find that pulling on the separate strands renders the undenatured state metastable at any temperature. If there is a transition, it is of the spinodal type \cite{spinodal}. Finally, in Section \ref{sec:conclusions} we review our analysis and point to its implications. 
%%%%%%%%%%%%%%%%%%%%%%%%%%%%%%%%%%%%%%%%%%%%%%%%%%%%%%%%%%%%%%%%%%%%%%%%%%%%%%%%%%%%%%%
%%%%%%%%%%%%%%%%%%%%%%%%%%%%%%%%%%%%%%%%%%%%%%%%%%%%%%%%%%%%%%%%%%%%%%%%%%%%%%%%%%%%%%%
%%%%%%%%%%%%%%%%%%%%%%%%%%%%%%%%%%%%%%%%%%%%%%%%%%%%%%%%%%%%%%%%%%%%%%%%%%%%%%%%%%%%%%%
\section{The effect of an external force on a freely-jointed chain} \label{sec:extforce}
%%%%%%%%%%%%%%%%%%%%%%%%%%%%%%%%%%%%%%%%%%%%%%%%%%%%%%%%%%%%%%%%%%%%%%%%%%%%%%%%%%%%%%%
%%%%%%%%%%%%%%%%%%%%%%%%%%%%%%%%%%%%%%%%%%%%%%%%%%%%%%%%%%%%%%%%%%%%%%%%%%%%%%%%%%%%%%%

\begin{figure}[htbp]
\begin{center}
\centerline{\includegraphics[width=2in]{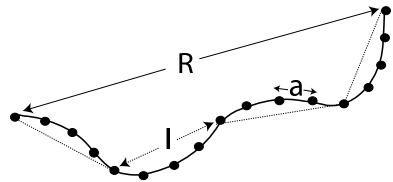}}
\caption{The Gaussian chain in which the monomeric units are separated by a distance $a$ can be modeled as a freely jointed chain with link length $l$. $l$ is the Kuhn length of the chain.}
\label{fig:Gausschain}
\end{center}
\end{figure}

For the purposes of calculating the effect of the external forces on a single strand of DNA, an appropriate starting point is the freely jointed chain (FJC).
A long segment of DNA can be approximated as a chain of many identical molecules connected with each other at joints which allow for spatial rotations (See Fig.~\ref{fig:Gausschain}).

The probability distribution of the end-to-end distance vector $\boldsymbol{R}$ of such an object is given by
\begin{eqnarray}
&P_N(\boldsymbol{R})=&\delta^{(3)}(\boldsymbol{R} -\sum_{n=1}^N   \boldsymbol{l}_n)\times\nonumber\\
&&\prod_{n=1}^N\left[ d^3 \boldsymbol{l}_n \,\frac{1}{4\pi l^2} \delta(|\boldsymbol{l}_n| -l)\right],
\label{eq:distribution}
\end{eqnarray} 
where $N$ is the number of units of the FJC, $l$ is the length of each unit. It can be shown that $l$ is equal to twice the persistence length of the chain \cite{Grosberg}.

The first delta  function on the right hand side of Eq.~(\ref{eq:distribution}) ensures that the vectors $\boldsymbol{l}_n$ of the chain elements add up to the distance vector $\boldsymbol{R}$.
  
If a force $\boldsymbol{F}$ is applied at one end of the chain and an equal and opposite force acts on the other end, then there is an additional Boltzmann factor weighting the chain configuration sum equal to $\exp[\boldsymbol{F}\cdot \boldsymbol{R}/k_BT]$ so that Eq.~(\ref{eq:distribution}) becomes

\begin{eqnarray}
&P_N(\boldsymbol{R},\boldsymbol{F})=&\exp\left[\frac{\boldsymbol{F} \cdot \boldsymbol{R}}{k_B T}\right]\delta^{(3)}(\boldsymbol{R} -\sum_{n=1}^N \boldsymbol{l}_n)\times\nonumber\\
&&\prod_{n=1}^N\left[d^3\boldsymbol{l}_n \,\frac{1}{4\pi l^2} \delta(|\boldsymbol{l}_n| -l)\right].
\label{eq:distr-F}
\end{eqnarray} 

Using the saddle point approximation, the distribution function in Eq.~(\ref{eq:distr-F}) can be evaluated as  
\begin{eqnarray}
&P_N(\boldsymbol{\rho},\boldsymbol{f})&\propto\left(\frac{\sinh f}{f}\right)^N\exp\left\{\frac{N}{2}\Phi^{(2)}_{\rho_{\perp}\rho_{\perp}}
\,\boldsymbol{\rho}_{\perp}^2\right\}
\times\nonumber\\
&&\exp\left\{\frac{N}{2}\Phi^{(2)}_{\rho_{\|}\rho_{\|}}\,(\rho_{\|}-\rho^*_{\|})^2\right\},  
\label{eq:distr_function}
\end{eqnarray}
where we have introduced the dimensionless parameters
\begin{eqnarray}
\boldsymbol{f}\equiv \frac{\boldsymbol{F}\, l}{k_B T},\qquad
\boldsymbol{\rho}=\frac{\boldsymbol R}{Nl}.
\end{eqnarray}
$f$ is the magnitude of $\boldsymbol{f}$; $\rho_{\|}$ and $\boldsymbol{\rho}_{\perp}$ are the components 
of $\boldsymbol\rho$ correspondingly parallel and perpendicular to the dimensionless force $\boldsymbol{f}$, and
\begin{equation}
\rho^*_{\|}=\coth f-\frac{1}{f}.
\end{equation}
The details of the calculation are presented in Appendix~\ref{sec:distribution}.
We have introduced the definitions of  the functions $\Phi^{(2)}_{\rho_{\perp}\rho_{\perp}}(f)$ and $\Phi^{(2)}_{\rho_{\|}\rho_{\|}}(f)$  in Eqs.~(\ref{eq:Phi1}) and (\ref{eq:Phi2}).
 
In the limit of weak forces, $f\ll 1$, 
\begin{eqnarray}
\Phi^{(2)}_{\rho_{\|}\rho_{\|}} \to   -3, \quad
\Phi^{(2)}_{\rho_{\perp}\rho_{\perp}} \to  -3, \quad 
\rho_{\|}^*  \to  \frac{f}{3},  \label{eq:limits}
\end{eqnarray}
and  the distribution function in Eq.~(\ref{eq:distr_function}) splits into a product of two terms, one describing the Gaussian chain and the other the statistical weight due to a small-force perturbation:
\begin{equation}
\label{weakf}
P_N(\boldsymbol{R},\boldsymbol{f})\propto \,\exp\Bigl\{-\frac{3 R^2}{2 N l^2}\Bigr\}
\exp\Bigl\{\boldsymbol{f}\cdot \boldsymbol{R}\Bigr\}.
\end{equation}
To evaluate the end-to-end separation of the chain under the influence of the external force we compute 
$\langle R^2 \rangle$,
\begin{eqnarray}
\lefteqn{\langle R^2 \rangle=\frac{\int (R_{\|}^2+\boldsymbol{R}_{\perp}^2)P_N(\boldsymbol{R}, \boldsymbol{f})dR_{\|}d\boldsymbol{R}_{\perp} }{\int P_N(\boldsymbol{R}, \boldsymbol{f}) dR_{\|}d\boldsymbol{R}_{\perp} } }\nonumber\\
&& = Nl^2 \left [ \frac{1}{(-\Phi^{(2)}_{\rho_{\|}\rho_{\|}})}+
\frac{2}{(-\Phi^{(2)}_{\rho_{\perp}\rho_{\perp}})}+N\,\rho^*_{\|}\right].
\label{end-to-end}
\end{eqnarray}
In the limiting case of weak forces $f\ll 1$ from Eq.~(\ref{end-to-end}), with the use of  Eq.~(\ref{eq:limits}), it follows 
\begin{equation}
\langle R^2 \rangle=Nl^2\left[\frac{1}{3}+\frac{2}{3}+0 \right]=Nl^2.
\end{equation}
Thus the DNA chain behaves as a Gaussian coil, with r.m.s. end-to-end distance proportional to $\sqrt{N}$.

\noindent For strong forces $f\gg 1$,  
\begin{equation}
\langle R^2 \rangle=Nl^2\left[0 + 0 + N \right]=(N\,l)^2,
\end{equation} 
that corresponds to a fully stretched chain.

We acquire more essential information by constructing the Fourier transformed generating function $\tilde{G}(z,\boldsymbol{q},\boldsymbol{f})$, defined formally in terms of a sum over {\it monomeric units} $k$ of the spatial Fourier transform of configurations.

We find
\begin{eqnarray}
\lefteqn{\tilde{G}(z,\boldsymbol{q}, \boldsymbol{f}) =\sum_{k=0}^{\infty}
z^k \tilde{P}_{k}(\boldsymbol{q},\boldsymbol{f})\nonumber}\\
&&=\sum_k z^k  
\int P_{k}(\boldsymbol{\rho},\boldsymbol{f})\, e^{i \boldsymbol{q}\cdot\boldsymbol{\rho}}\,d^3\rho\nonumber\\
&& \propto\sum_{k} \,\left(\frac{z}{z_0}\right)^k\,e^{k (- X_{\|}\,q^2_{\|}
-X_{\perp}\,\boldsymbol{q}^2_{\perp}+i\,Y_{\|}\,q_{\|})}\label{eq:q_propagator}\\
&&=\left[1- \frac{z}{z_0}\,e^{- X_{\|}\,q^2_{\|} -X_{\perp}\,\boldsymbol{q}^2_{\perp}+ i\,Y_{\|}\,q_{\|}} \right]^{-1}
\nonumber\\
&&  
\simeq\left[1- z/z_0 + X_{\|}\, q^2_{\|} 
+ X_{\perp}\,\boldsymbol{q}^2_{\perp} - i\,Y_{\|}\,q_{\|} \right]^{-1},\nonumber
\end{eqnarray}
where 
\begin{equation}
\label{eq:z0}
z_0=\Bigl(\frac{f}{\sinh f}\Bigr)^{a/l}
\end{equation}
is the critical fugacity,
with
\begin{eqnarray}
&X_{\|}(f)&=- la/(2\Phi^{(2)}_{\rho_{\|}\rho_{\|}})>0 \\
&X_{\perp}(f)&=-la/(2\Phi^{(2)}_{\rho_{\perp}\rho_{\perp}})>0\\
&Y_{\|}(f)&=a \rho_{\|}^*>0,
\end{eqnarray}
and $a$ is the size of a monomeric unit.
 
The last line of Eq.~(\ref{eq:q_propagator}) reflects the fact that we are interested in the behavior of the generating function when $z \approx z_0$ and in the limit of small $\boldsymbol{q}$. 

\begin{figure}[htbp]
\begin{center}
\includegraphics[width=3in]{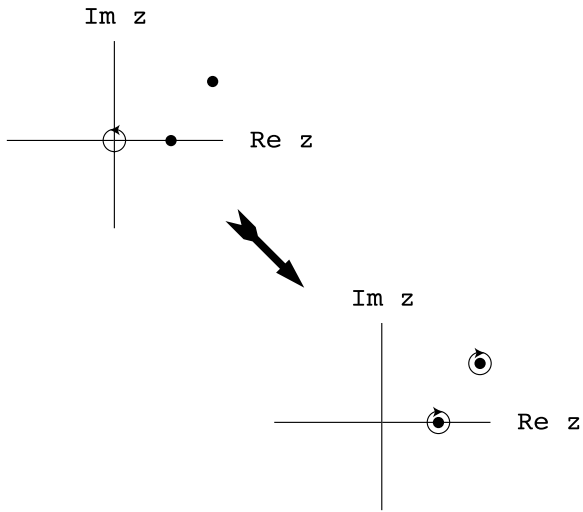}
\caption{The contours utilized in the evaluation of the integral Eq.~(\ref{eq:Gauss4}) leading to the extraction of the $k^{\rm th}$ power of $z$ in the expansion of the generating function. On the upper left hand side: the original contour, consistent with the extraction of that coefficient via Cauchy's theorem. On the lower right hand side: the distortion of the original contour to enclose singularities of the integrand---in the illustrated case simple poles---located at the heavy dots in the figure.}
\label{fig:contour}
\end{center}
\end{figure}

We will extract from the above generating function the number of weighted configurations of a $k$-monomer chain. We do this by calculating the coefficient of $z^k$ in the power series expansion of the generating function, which we take to be given by the last line of Eq.~(\ref{eq:q_propagator}). The actual calculation makes use of Cauchy's theorem \cite{Jeffreys} and  involves the following contour integration:
\begin{equation}
\frac{1}{2 \pi i}\oint \frac{\tilde{G}(\,z, \boldsymbol{q} =0, \boldsymbol{f}\,)}{z^{k+1}} dz
\label{eq:Gauss4}
\end{equation}
where the contour is as illustrated on the upper left hand side of Fig.~\ref{fig:contour}.
The evaluation of the integral involves the distortion of the integration contour, as indicated in the lower right hand side of Fig.~\ref{fig:contour}, so as to enclose the singularities in the generating function. The figure illustrates the result of that distortion when the singularities are two simple poles. In our case, the generating function possesses a single pole, which lies on the real axis. When there is more than one singularity, the dominant contribution arises from the singularity that lies closest to the origin. In fact, in the thermodynamic limit, $k \rightarrow \infty$, effectively the only contribution that matters is the one generated by the singularity closest to $z=0$. This is not an issue in the calculation performed here, but it will be as we consider the mathematics of melting as embodied in the Poland Scheraga model and modifications thereof. 

Continuing, we note, as indicated immediately above, that the contour integration is dominated by the single contribution at the pole in  the function on the last line of Eq.~(\ref{eq:Gauss4}), with $\boldsymbol{q}=0$,  corresponding to the solution of the equation 
\begin{equation}
 1-z/z_0 =0.
\label{eq:Gauss5}
\end{equation}
Inserting this solution into the result of the contour integration illustrated in the lower right hand corner of Fig.~\ref{fig:contour}, and making note of the residue, we end up with the following result for the total number of weighted $k$-monomer configurations when the polymer is subjected to the externally generated tension $\boldsymbol{f}$
\begin{equation}
\frac{1}{2 \pi i}\oint \frac{\tilde{G}(\,z, \boldsymbol{q} = 0, \boldsymbol{f}\,)}{z^{k+1}} dz \ \to\  \left(\frac{\sinh f }{f} \right)^{ka/l}.
\label{eq:Gauss6}
\end{equation}

%%%%%%%%%%%%%%%%%%%%%%%%%%%%%%%%%%%%%%%%%%%%%%%%%%%%%%%%%%%%%%%%
\subsection{Corrections for excluded volume: one loop order}
%%%%%%%%%%%%%%%%%%%%%%%%%%%%%%%%%%%%%%%%%%%%%%%%%%%%%%%%%%%%%%%

\begin{figure}[htbp]
\begin{center}
\includegraphics[width=1.5in]{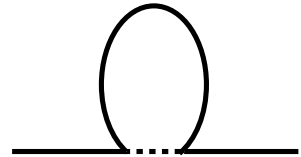}
\caption{The first order correction to the generating function arising from excluded volume. The dashed line is the effective repulsive interaction arising from self-avoidance.}
\label{fig:oneloop}
\end{center}
\end{figure}

We can now assess the interplay of this externally-generated tension and self-avoidance in influencing the asymptotic statistics of an excluded volume Gaussian polymer. A way to assess this interplay is to consider the lowest order correction to the generating function, as shown in Fig.~\ref{fig:oneloop}.

The one-loop correction corresponds to the expression
\begin{eqnarray}
\lefteqn{u \int \tilde{G}(z,\boldsymbol{q}, \boldsymbol{f}\,\,) dq_{\|}d\boldsymbol{q}_{\perp} \nonumber} \\ 
& = & u \int \left[1-z/z_0 + X_{\|} q^2_{\|} + X_{\perp}\boldsymbol{q}_{\perp}^2-
i Y_{\|}q_{\|}   \right]^{-1}\! dq_{\|}d\boldsymbol{q}_{\perp} \nonumber \\
& \propto &u \int \frac{d^3q}{1-z/z_0 + Y^2_{\|}/4X_{\|}+\boldsymbol{q}^2} \nonumber \\
& \propto & u \left( A(z) - \sqrt{1 - z/z_0 + Y^2_{\|}/4 X_{\|}  } \,\, \right),
\label{eq:Gauss7}
\end{eqnarray}
where $u$ is the coupling constant measuring the strength of the repulsive interaction. $A(z)$ is a singularity-free function of $z$. The next-to-last line in Eq.~(\ref{eq:Gauss7}) follows from a shift in the contour of integration over the component of $\boldsymbol{q}$ parallel to $\boldsymbol{f}$.

The outcome of this calculation is that the one-loop correction to the effective self-energy of the generating function of the excluded volume yields an expression having the following form
\begin{eqnarray}
\lefteqn{\tilde{G}(z,\boldsymbol{q}, \boldsymbol{f}) = \bigg [1-z/z_0 + X_{\|} q^2_{\|} + X_{\perp}\boldsymbol{q}_{\perp}^2 -
i Y_{\|}q_{\|} \nonumber }\\
&&{} + u \,A(z) - u\sqrt{1 - z/z_0 + Y^2_{\|}/4 X_{\|}}\,\, \bigg ]^{-1}.
\label{eq:Gauss8}
\end{eqnarray}

Setting $\boldsymbol{q}=0$, in order to locate the singularity that dominates a calculation of the total number of weighted configurations, we find
\begin{eqnarray}
\lefteqn{\tilde{G}(z,\boldsymbol{q}=0, \boldsymbol{f}\,) = \bigg [\, 1- z/z_0 + u \,A(z) \nonumber} \\ 
&& { }-u\sqrt{1- z/z_0 + Y^2_{\|}/4 X_{\|}}\,\, \bigg ]^{-1}.
\label{eq:Gauss9}
\end{eqnarray}

The singularity in the function, if $u$ is sufficiently small, is slightly shifted from the unrestricted value.

However, the singularity of the one-loop contribution is given by

\begin{equation}
z_c = z_0\bigg(1 + Y^2_{\|}/4X_{\|}\bigg).
\label{eq:zpole}
\end{equation}

In the weak-force limit it is equal to
\begin{equation} 
\label{eq:zc}
z_c = z_0\left(1+\frac{1}{6}\frac{a}{l}f^2\right).
\end{equation}

\begin{figure}[htbp]
\begin{center}
\includegraphics[width=2.5in]{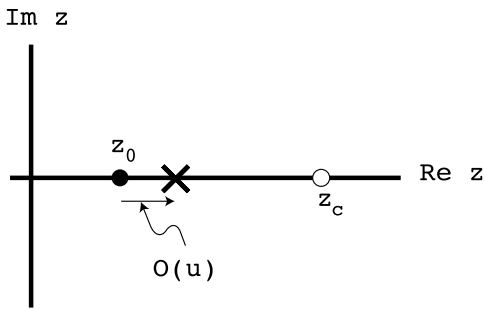}
\caption{At lowest nontrivial loop order, the original singularity in the generating function
for $\boldsymbol{q}=0$, at $z=z_0$, (black circle) is shifted a distance $O(u)$ to the right (cross), and lies closer to the origin than the singularity in the one-loop correction at $z=z_c$ (open circle). }
\label{fig:shift}
\end{center}
\end{figure}

In particular, the singularity in the generating function remains closer to the origin than the singularity in the one-loop contribution in Eq.~(\ref{eq:Gauss7}). This means that the latter singularity will not have an effect on the asymptotic statistics of the polymer chain under tension, in contrast to the situation of the unstretched chain, for which the one-loop correction for self-avoidance exerts an essential, and transforming, influence on asymptotic configurational statistics. The shift is indicated in Fig.~\ref{fig:shift}.

%%%%%%%%%%%%%%%%%%%%%%%%%%%%%%%%%%%%%%%%%%%%%%%%%%%%%%%%%%%%%%%%%%%%%%%%%%%
\subsection{Two loop order correction} \label{sec:twoloop}
%%%%%%%%%%%%%%%%%%%%%%%%%%%%%%%%%%%%%%%%%%%%%%%%%%%%%%%%%%%%%%%%%%%%%%%%%%%

The next obvious question is whether this difference between the singularity structure of the corrected generating function and the singularities of the corrections persists to higher loop order. A cursory investigation of the general structure of such contributions to asymptotic configurational statistics indicates that, if the strength of the self-avoiding interaction is low enough, the singularity structure of higher loop order corrections will not lead to an essential renormalization of loop statistics.

As the argument utilized here is based on the structure of excluded volume corrections in real space, we begin with the form of the generating function in real space, the inverse Fourier transform of $\tilde{G}(z,q_{\|},\boldsymbol{q}_{\perp}, f)$. We find for this quantity

\begin{eqnarray}
\lefteqn{G(z,\boldsymbol{R},\boldsymbol{f})\nonumber}\\
&&{}\propto \int\frac{e^{-i(\,q_{\|}R_{\|}+\,\boldsymbol{q}_{\perp}\boldsymbol{R}_{\perp})} \,dq_{\|}d\boldsymbol{q}_{\perp}} {1- z/z_0 + X_{\|}\, q^2_{\|} +X_{\perp}\,\boldsymbol{q}^2_{\perp} - i\,Y_{\|}\,q_{\|} }\label{eq:rs1} \\
&&{}\propto\, 
\frac{1}{r} \exp\left\{- r\sqrt{1-\frac{z}{z_0} +\frac{Y^2_{\|}}{4 X_{\|}}}+\frac{Y_{\|}\,z}{2\,\sqrt{X_{\|}}}
\right\},
\nonumber
\end{eqnarray}
 where we have introduced a new variable
\begin{equation}
\boldsymbol{r}=(\boldsymbol{r}_{\perp}, z)=\left(\frac{\boldsymbol{R}_{\perp}}{\sqrt{X_{\perp}}}, \frac{R_{\|}}{\sqrt{X_{\|}}} \right).
\end{equation}

In particular, in the limit of weak forces, $f\ll 1$, we find in the frist order of $f$
\begin{equation}
X_{\|}=X_{\perp}=la/6, \qquad Y_{\|}=af/3,
\end{equation}
and
\begin{equation}
\label{eq:Gweakf}
G(z,\boldsymbol{r},\boldsymbol{f})\propto \frac{e^{f\,r\cos\theta/l}e^{-r \sqrt{1-z/z_0+f^2 a/6l}\sqrt{6/al}}}{r},
\end{equation}
where we assumed that the force in the $\hat{z}$ direction.

The real space version of the one-loop self energy shown in Fig.~\ref{fig:oneloop} can be reproduced to within non-singular contributions by allowing $\boldsymbol{r}$ to become very small in Eq.~(\ref{eq:Gweakf}), looking in particular at the zeroth order in $r$ terms in the resulting expression. Expanding the result in powers of $\boldsymbol{r}$,  we find for the one-loop correction
\begin{equation}
\frac{1}{r}+ \sqrt{6/al} \sqrt{1- z/z_c} + \frac{f \cos \theta}{l} +O\left(r\right),
\label{eq:oneloop1}
\end{equation}
where $z_c$ is given by Eq.~(\ref{eq:zc}).

Note that the next to last term in Eq.~(\ref{eq:oneloop1}) depends on the direction in which the displacement vector points. This contribution is, in fact, a singular one, in that it is indeterminate in the limit $\boldsymbol{r} =0$. However, the averaging inherent in compensating for self-intersection is over directions, which means that we must take the angular average $\langle \cos \theta \rangle$, which is equal to zero. We thus recover the singular portion of the one-loop correction as proportional to $\sqrt{1-z/z_c}$, as we did in the previous section. 

\begin{figure}[htbp]
\begin{center}
\includegraphics[width=2.5in]{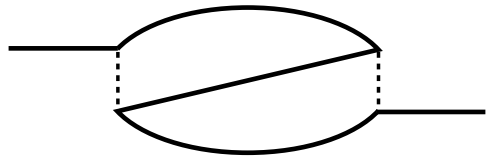}
\caption{The two loop insertion.}
\label{fig:twoloop}
\end{center}
\end{figure}

We now turn to the two loop self energy. Fig.~\ref{fig:twoloop} illustrates the most interesting two loop term. 
The structure of this figure is fairly straightforward. It has the following spatial Fourier transform
\begin{eqnarray}
u^2 \int e^{i\boldsymbol{q}\cdot\boldsymbol{r}}G(z,\boldsymbol{r},\boldsymbol{f}=0)^3 d^3r. 
\label{eq:2order}
\end{eqnarray}

As the dominant issue for us is the way in which this correction affects the total number of configurations, we are interested in the $\boldsymbol{q}=0$ limit of the above expression. Inserting   Eq.~(\ref{eq:Gweakf})  into Eq.~(\ref{eq:2order}), with $f$ set equal to zero, we obtain the following result for the two loop self energy

\begin{equation}
\propto u^2
\int
\bigg(\frac{e^{-r \sqrt{1-z/z_0+f^2 a/6l}\sqrt{6/al}}}{r}\bigg )^3 d^3r.
\label{eq:tlnf}
\end{equation}

When there is tension, the integrand in Eq.~(\ref{eq:tlnf}) is replaced by the one that contains a Boltzmann factor associated with the applied forces. The integral to perform in this case is

\begin{equation}
\propto u^2
\int d\Omega \int_{r_0}^{\infty} r^2 dr
\bigg(\frac{e^{-r \sqrt{1-z/z_0+f^2 a/6l}\sqrt{6/al}}}{r}\bigg )^3 e^{fz/l}.  
\label{eq:tlnf2}
\end{equation}

Carrying out the integrations, we end up with the following result for the two loop self energy associated with the diagram in Fig.~\ref{fig:twoloop}.
%\begin{widetext}
\begin{eqnarray}
&&\propto u^2 \frac{1}{r_0}
\left\{ \mathcal{G}
\left[\left(3 \sqrt{6/al} \sqrt{1-z/z_0+f^2 a/6l} -f/l \right)r_0\right]\right.\nonumber \\
&&\left. {}- \mathcal{G}
\left[\left( 3\sqrt{6/al}\sqrt{1-z/z_0+f^2 a/6l} + f /l \right)r_0\right] \right\},
\label{eq:twoloop1}
\end{eqnarray}
%\end{widetext}
where 
\begin{equation}
\mathcal{G}(x) = e^{-x} + x \mathop{\rm Ei}(-x).
\label{eq:twoloop2}
\end{equation}
Taking the $r_{0}\rightarrow 0$ limit of Eq.~(\ref{eq:twoloop1}), we find that there is a singularity of the form $w \ln w$, where 
\begin{equation}
w = 3 \sqrt{6/al} \sqrt{1-z/z_0+f^2 a/6l} \pm f/l.
\label{eq:twoloop3}
\end{equation}
This tells us that the leading order singularity in the two loop self energy lies on the following location on the real $z$ axis:
\begin{equation}
z=z_0(1+4 f^2 a/27 l ).
\label{eq:twoloop4}
\end{equation}
This is to the left of $z_c=z_0(1 + f^2 a/6 l)$, but further from the origin than $z_0$, the leading singularity of the unrestricted linear chain polymer under the influence of tension, as given by Eq.~(\ref{eq:z0}). 

A simple---but we believe essentially correct---argument holds that the energy penalty associated with the necessity of looping back against the tension that is required for self-intersection militates against a fundamental alteration of configurational statistics when corrections are made for excluded volume.  For a more extended discussion, see Appendix~ \ref{sec:higherorder}.

%%%%%%%%%%%%%%%%%%%%%%%%%%%%%%%%%%%%%%%%%%%%%%%%%%%%%%%%%%%%%%%%%%%%%%%%%%%%%%%%%
\section{Effect of self-avoidance on the denaturation bubble} \label{sec:bubble}
%%%%%%%%%%%%%%%%%%%%%%%%%%%%%%%%%%%%%%%%%%%%%%%%%%%%%%%%%%%%%%%%%%%%%%%%%%%%%%%%%

\begin{figure}[htbp]
\begin{center}
\includegraphics[width=3in]{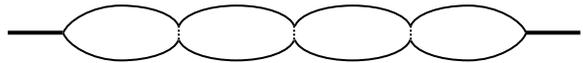}
\caption{The denaturation bubble, with corrections for self-interaction incorporated. Dotted lines show the repulsive interactions between the strands.}
\label{fig:ladder}
\end{center}
\end{figure}

In the case of the melting DNA chain, there is one way in which self avoidance exerts a fundamental influence on the statistics, and thereby the thermodynamics, of the system. This is through the modification of the contribution of the denaturation bubble to the partition function of the system. As noted previously, configurations in which self-intersection requires that the chain loop back on itself, in opposition to the imposed tension, are, by reason of the energy penalty associated with such a configuration, rendered irrelevant to the asymptotic statistics of thermal denaturation. This means that the vertex correction found by Kafri {\em et al.} \cite{kmp} to cause melting to become first order in the absence of tension  plays no such transformative role when that tension is present. However, intersections of the two strands in the denaturation bubble as shown in Fig.~\ref{fig:ladder} must be taken into account in the evaluation of the partition function sum of the melting chain. 

In the denaturation bubble there are equal number of units in its lower and upper chains. To enforce this condition in the context of the calculation that we perform below, we assign different fugacites, $z_{1,2}$ to the units in the two strands in the bubble. The relation between resulting generating function of the, in general, asymmetric loop ${\cal F}(z_1,z_2)$and the generating function representing the denatured loop is the following 
 \begin{eqnarray}
\nonumber
&\displaystyle\Pi(z)=\frac{1}{2\pi i} \oint \frac{dw}{w} {\cal F}(w\sqrt{z},\frac{\sqrt{z}}{w}).&\\
\label{loopGF}
\end{eqnarray}
where the contour of integration is around the origin.
Indeed, by definition,
\begin{equation}
\displaystyle{\cal F}(z_1,z_2)=\sum_m\sum_n C_{mn} z_1^m z_2^n,
\end{equation}
where $C_{nm}$ is the partition function of the loop formed by two chains of $n$ and $m$ units correspondingly.

Then according to Eq.~(\ref{loopGF}), the generating function of the denaturation bubble becomes
\begin{eqnarray}
&\Pi(z)&=\frac{1}{2\pi i}\sum_m\sum_n C_{mn}\oint_C\frac{dw}{w} (w\sqrt{z})^m \left(\frac{\sqrt{z}}{w } \right)^n\nonumber\\
&&=\sum_m\sum_n C_{mn} z^{m/2}z^{n/2}\frac{1}{2\pi}\int_0^{2\pi} d\phi
e^{i(m-n)\phi}\nonumber\\
&&=\sum_m\sum_n C_{mn} z^{m/2}z^{n/2}\delta_{mn}=\sum_n C_{nn} z^n,
\label{eq:symmetric-loop}
\end{eqnarray}
that shows that $\Pi(z)$ is indeed the generating function of a bubble with equal number of units in each chain.
In the second line of Eq.~(\ref{eq:symmetric-loop}) we have deformed the contour of integration to a unit circle with the center at the origin.
To account the possible interactions of the two chains we need to perform a Dyson-like summation over the loops formed by the strands interactions as shown in Fig.\ref{fig:ladder}. Thus the generating function for the self-interacting denaturation bubble becomes
\begin{eqnarray}
&\displaystyle\Pi(z)=\frac{1}{2\pi i} \oint \frac{dw}{w} \sum_n^{\infty}(-u)^n{\cal F}^n(w\sqrt{z},\frac{\sqrt{z}}{w}).&
\label{eq:Pi}
\end{eqnarray}
In the next section we derive the closed-form expression for $\Pi(z)$.
 
%%%%%%%%%%%%%%%%%%%%%%%%%%%%%%%%%%%%%%%%%%%%%%%%%%%%%%%%%%%%%%%%%%%%%%%%%%%%%%%%%%%%%%%%%%%%%%%% 
\subsection{Mutual avoidance of the two chains in a loop --- performing the  
ladder sum} \label{sec:mutualavoidance}
%%%%%%%%%%%%%%%%%%%%%%%%%%%%%%%%%%%%%%%%%%%%%%%%%%%%%%%%%%%%%%%%%%%%%%%%%%%%%%%%%%%%%%%%%%%%%%%% 
The denaturation bubble under the action of the external force $f$ behaves as a set of two parallel springs, each spring under stress $f/2$.
To find the generating function of the loop we form the integral
\begin{eqnarray}
&{\cal F}(z_1,z_2)&\int dR_{\|} d\boldsymbol{R}_{\perp}G(z_1,R_{\|},\boldsymbol{R}_{\perp},f/2)\times\nonumber\\
&&G(z_2,R_{\|},\boldsymbol{R}_{\perp},f/2).    
\label{eq:Fz1z2} 
\end{eqnarray}
Plugging  the last line of Eq.~(\ref{eq:rs1}) into Eq.~(\ref{eq:Fz1z2}) yields 
\begin{eqnarray}
\nonumber
&{\cal F}(z_1,z_2)\propto\int d^3 r
\frac{1}{r^2}\exp \left\{-r\sqrt{1-z_1/\tilde{z}_0+y_{\|}^2/4x_{\|}}\right.&\\
&\left. {}-r\sqrt{1-z_2/\tilde{z}_0+ y_{\|}^2/4x_{\|}}+z y_{\|}/\sqrt{x_{\|}} \, \right\}
\label{eq:intFz1z2}& 
\end{eqnarray}
with
\begin{equation}
x_{\|}(f)\equiv X_{\|}(f/2),\
y_{\|}(f)\equiv Y_{\|}(f/2),\
\tilde{z}_0\equiv z_0(f/2).
\end{equation}

The angular integral in Eq.~(\ref{eq:intFz1z2}) leads to 
\begin{equation}
\label{eq:trick1}
\frac{e^{-a_1 r}-e^{-a_2 r}}{r^3}, 
\end{equation}
where we have introduced the following notations
\begin{eqnarray}
&a_{1,2}&=\sqrt{1-z_1/\tilde{z}_0+y_{\|}^2/4x_{\|}}\nonumber\\
&&{}+\sqrt{1-z_2/\tilde{z}_0+y_{\|}^2/4x_{\|}}\mp y_{\|}/\sqrt{x_{\|}}. 
\end{eqnarray}
Multiplying  Eq.~(\ref{eq:trick1}) by $r^2$ and integrating from $r=0$ to $\infty$ we obtain
\begin{equation}
\label{eq:trick2}
\int_0^{\infty}\frac{e^{-a_1 r}-e^{-a_2 r}}{r} dr.
\end{equation}
The integral in Eq.~(\ref{eq:trick2}) can be rewritten as 
\begin{equation}
\int_0^{\infty} dr \left\{\int_{a_1}^{a_2} e^{-w r} dw \right\}.
\end{equation}
Exchanging the orders of integration, we are left with
\begin{equation}
\ln\frac{a_2}{a_1}.
\end{equation}
Applying this procedure to the integral in Eq.~(\ref{eq:intFz1z2}), we obtain the following expression
\begin{widetext}
\begin{eqnarray}
&&{\cal F}(z_1,z_2)\propto
\frac{\sqrt{x_{\|}} }{y_{\|}} 
\ln\left[\frac{\sqrt{1-z_1/\tilde{z}_0 + y_{\|}^2/4 x_{\|}}+\sqrt{1-z_2/\tilde{z}_0 + y_{\|}^2/4 x_{\|}} + y_{\|}/\sqrt{x_{\|}}}{ \sqrt{1-z_1/\tilde{z}_0 + y_{\|}^2/4 x_{\|}}+\sqrt{1-z_2/\tilde{z}_0 + y_{\|}^2/4 x_{\|}} -   y_{\|}/\sqrt{x_{\|}}}\right].
\label{eq:loop-z1z2}  
\end{eqnarray}
\end{widetext}
Plugging $z_1=\sqrt{z}e^{i\theta}$ and $z_2=\sqrt{z}e^{-i\theta}$ into Eq.~(\ref{eq:loop-z1z2})
and expanding in power series of $\theta$ we obtain for
\begin{eqnarray}
&&\label{eq:sqrt}
\Big[\sqrt{1-\sqrt{z}e^{i\theta}/\tilde{z}_0 +y_{\|}^2/4 x_{\|}}+\mbox{c.c.}\Big] - y_{\|}/\sqrt{x_{\|}}=\\
&&\nonumber \Big[\sqrt{1-\sqrt{z}/\tilde{z}_0 + y_{\|}^2/4 x_{\|} }\,(1-i\theta-\frac{\theta^2}{2})+\mbox{c.c.}\Big]-
\frac{y_{\|}}{\sqrt{x_{\|}}},
\end{eqnarray}
where ``c.c.'' stands for ``complex conjugate.'' In the second line of Eq.~(\ref{eq:sqrt}) we have rescaled the variable $\theta$.

Denoting $\delta=1-z/\tilde{z}_0^2$ and expanding Eq.~(\ref{eq:sqrt}) with respect to $\delta$ we end up with

\begin{equation}
\Big[y_{\|}/2\sqrt{x_{\|}} + \delta -i\theta + \frac{\theta^2}{2}+\mbox{c.c.}\Big]-\frac{y_{\|}}{\sqrt{x_{\|}}} \propto \delta + \theta^2,
\label{eq:expansion}
\end{equation}
where, once again, we have rescaled the variable $\theta$.

%\begin{eqnarray}
%\lefteqn{{\cal F}(z_1,z_2)\propto
%\frac{\sqrt{x_{\|} } }{y_{\|}} 
%\ln\Big[\sqrt{1-z_1/\tilde{z}_0+y_{\|}^2/4x_{\|}}  }\nonumber\\
%&&{}+\sqrt{1-z_2/\tilde{z}_0+y_{\|}^2/4x_{\|}}+ y_{\|}/\sqrt{x_{\|}}\Big]\nonumber\\
%&&{}-\ln\Big[\sqrt{1-z_1/\tilde{z}_0+y_{\|}^2/4x_{\|}}\\
%&&{}+\sqrt{1-z_2/\tilde{z}_0+y_{\|}^2/4x_{\|}}-y_{\|}/\sqrt{x_{\|}}\Big]
%\end{eqnarray}

%The expression for the generation function of the loop in the absence of the force, $f\to0$, is 
%given by
%$$
%{\cal F}(z_1,z_2)\propto \frac{1}{\sqrt{1 - z_1} + \sqrt{1 - z_2}}
%$$

Performing the geometrical sum in Eq.~(\ref{eq:Pi}) and using the results of the expansion in Eq.~(\ref{eq:expansion}) we obtain the following expression for the denaturation bubble generating function 
\begin{equation}
\int_0^{\theta_0} \frac{d \theta}{1 + A u - u\ln \left( \delta + \theta^2 \right)}.
\label{eq:intsum1}
\end{equation}
In Eq.~(\ref{eq:intsum1}) we have deformed the contour of integration to a unit cirlce and have assumed that the dominant contribution to the intergral comes from a small section of the contour between the pole of the integrand in Eq.~(\ref{eq:intsum1}) and the branch point at 
\begin{equation}
\label{eq:fugacity-ss}
\tilde{z}^2_0=\left(\frac{f/2}{\sinh f/2}\right)^{2a/l}.
\end{equation}  
A route to the determination of the behavior of this integral when $\delta$ is small is to take the derivative of the integral in Eq.~(\ref{eq:intsum1}) with respect to $\delta$. Once the integral that results is evaluated,  one integrates with respect to $\delta$ to reconstruct the integral of interest. The $\delta$-derivative is straightforward and results in
\begin{equation}
\int_0^{\theta_0} \frac{d \theta}{\left(1 + A u - u \ln \left(\delta + \theta^2\right)\right)^2}\frac{u}{\delta + \theta^2}.
\label{eq:int1}
\end{equation}
In the case of this integral, we can extend the upper limit of integration to infinity without encountering a divergence. Replacing the integration variable $\theta$ by $y = \theta/ \delta$, we are left with the following integral:
\begin{equation}
\frac{u}{\sqrt{\delta}}\int_0^{\infty} \frac{dy}{\left(1 + A u - u \ln\delta - u \ln \left(1 + y^2\right)\right)^2}\frac{1}{1+y^2}.
\label{eq:int2}
\end{equation}
When $|u \ln \delta| \gg 1$, the dominant contribution to the integrand in Eq.~(\ref{eq:int2}) is
\begin{eqnarray}
\lefteqn{\frac{u}{\sqrt{\delta}}\int_0^{\infty} \frac{dy}{\left( 1+A u - u \ln \delta \right)^2}\frac{1}{1+y^2}} \nonumber \\ & = &  \frac{\pi}{2} \frac{u}{\sqrt{ \delta}} \frac{1}{\left(1 + A u - u \ln \delta \right)^2} \nonumber \\
& \rightarrow & \frac{\pi}{2} \frac{u}{\sqrt{ \delta}} \frac{1}{u^2 \ln^2 \delta}.
\label{eq:int3}
\end{eqnarray}
It is now possible to integrate the expression above with respect to $\delta$. This leads to the following result
\begin{equation}
\frac{1}{u} \left( \mathop{\rm Ei} \left(\frac{\ln \delta}{2} \right) - \frac{\sqrt{\delta}}{\ln \delta }\right)
\frac{\pi}{2},
\label{eq:int4}
\end{equation}
where Ei is an exponential integral. Expanding the above result with respect to $\delta$ when that quantity is small, we find
\begin{equation}
\frac{\pi}{2 u} \left( \frac{8}{{\log (\delta )}^3} + 
  \frac{2}{{\log (\delta )}^2} \right) \sqrt{\delta},
  \label{eq:int5}
\end{equation}
which tells us that as $\delta \rightarrow 0$, the ladder diagram sum will be dominated by terms going as
\begin{equation}
\frac{\sqrt{\delta}}{( \ln \delta)^2}.
\label{eq:finalint}
\end{equation}
Equation (\ref{eq:finalint}) tell us that self-avoidance leads to a logarithmic modification of the contribution of the ``unrestricted'' denaturation bubble. 
%%%%%%%%%%%%%%%%%%%%%%%%%%%%%%%%%%%%%%%%%%%%%%%%%%%%%%%%%%%%%%%%%%%%
\section{The net effect of stress on the melting transition.}
\label{sec:effects}
%%%%%%%%%%%%%%%%%%%%%%%%%%%%%%%%%%%%%%%%%%%%%%%%%%%%%%%%%%%%%%%%%%%%

According to the Poland-Scheraga model \cite{ps1,ps2}, one can construct the grand partition function of the denaturing chain by taking the geometric sum of sequences of intact portions and denatured bubbles. If we call the grand partition function of an intact chain $G(z)$ and the grand partition function of a bubble $\Pi(z)$, then the overall grand partition function is 

\begin{eqnarray}
&G(z) + G(z)\Pi(z)G(z) + G(z)\Pi(z)G(z)\Pi(z)G(z) + ... &\nonumber\\
&\displaystyle=\frac{G(z)}{1 - G(z)\Pi(z)}= \frac{1}{G(z)^{-1} - \Pi(z)}.&
\label{eq:geomsum}
\end{eqnarray}

We model both double stranded (ds) and single stranded (ss) chains of DNA as freely jointed chains with different unit lengths, to take into account the greater flexibility of a ss chain.

The grand partition function of the double stranded segment of the molecule is 

\begin{eqnarray}
&G(z)=&\sum_k \left(\frac{z}{z_{ds}}\right)^k =\left(1-\frac{z}{z_{ds}}\right)^{-1},
\label{eq:dsgenfun}
\end{eqnarray}
where the summation runs over monomeric units.
The critical fugacity of the ds chain $z_{ds}$ in Eq.~(\ref{eq:dsgenfun}) is given by 
\begin{equation}
z_{ds}(T,F)=e^{-\Delta g/k_B T} \left(\frac{Fl_{ds}/k_B T}{\sinh Fl_{ds}/k_B T}\right)^{d_{ds}/l_{ds}}.
\label{eq:ds-fugasity}
\end{equation}
where $d_{ds}$ is the distance between adjacent base pairs in ds segment, $l_{ds}$ is the Kuhn  length of the ds segment.
$\Delta g=g_{ss}-g_{ds}$ is the difference of Gibbs free energies {\it per base pair}. 
The second factor in Eq.~(\ref{eq:ds-fugasity}) is associated with the configurational entropy of the ds chain  (ref. Eq.~(\ref{eq:Gauss6})).

For the generating function of the denaturation bubble, according to Eq.~(\ref{eq:finalint}), we have

\begin{equation}
\label{eq:Pi-ss}
\Pi(z)=\frac{\sqrt{1-z/z_{ss}}}{\Bigl[\log(1-z/z_{ss})\Bigr]^2},
\end{equation}
where $z_{ss}$ is the critical fugacity of ss strand. From Eq.~(\ref{eq:fugacity-ss}) it follows that
\begin{equation}
\label{eq:z-ss}
z_{ss}(T,F)=\left(\frac{F l_{ss}/2 k_B T}{\sinh F l_{ss}/2 k_B T}\right)^{2 d_{ss}/l_{ss}},
\end{equation}
where, $l_{ss}$ is the Kuhn length of a ss strand, $d_{ss}$ is the distance between adjacent base pairs in a ss segment. 

\begin{figure}[htbp]
\begin{center}
\centerline{\includegraphics[width=2.7in]{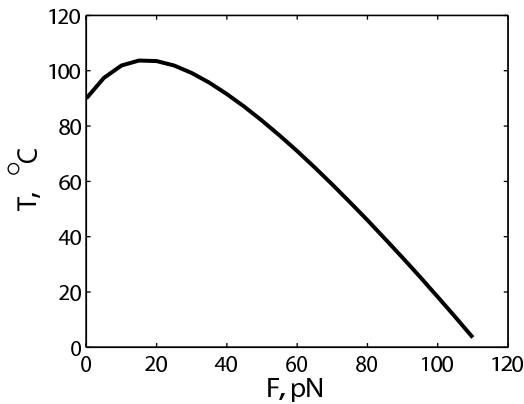}}
\caption{The phase diagram: critical temperature vs the applied force F. The following parameters were used:
distance between adjacent base pairs of ss chain $d_{ss}=0.58$nm, and of ds chain, $d_{ds}=0.34$ nm; persistent length of a ss chain 0.7nm, of a ds chain 50nm, $\Delta g=\Delta h -T\Delta s$ with
$\Delta s =12.5 k_B$, $\Delta h=T_c\Delta s$, $T_c=360$K \cite{RandB1}. }
\label{fig:phase}
\end{center}
\end{figure}

Strand separation occurs when the simple pole of the generating function of the DNA chain, Eq.~(\ref{eq:geomsum}), coincides with the branch point of $\Pi(z)$ function, 
which results in the relation
\begin{equation}
\label{phase_diagram}
z_{ds}(T,F)=z_{ss}(T,F).
\end{equation}
Solving Eq.~(\ref{phase_diagram}) numerically, we generate the phase diagram curve displayed in Fig. \ref{fig:phase}.  
An interesting feature of the phase diagram, as was also  discussed in \cite{RandB1}, is the presence of a turning point where $dT/dF$ changes sign.

The relation between the strain and stress for the freely jointed chain of link length $l$ is given by \cite{Grosberg}
\begin{equation}
\frac{\Delta L }{L}=\frac{|{\bf R }|}{Nl}=\coth x-\frac{1}{x}, \qquad x=\frac{Fl}{k_B T}.
\end{equation}
In the limit of the weak forces the chain's response to stress is linear, with effective spring constant
\begin{equation}
k_{ef}\propto \frac{k_B T}{ N l^2}.
\end{equation}
Since this spring constant in inversely proportional to the chain's length, a weak force aligns longer chains more easily than shorter ones. Thus the double stranded state of the DNA is more favorable than the denatured one in the weak force limit. As the force increases in strength the difference of stretching per base pair of ds and ss chains decreases and becomes negative. This is the point where $dT/dF$ changes sign. At large forces, when the molecule is stretched nearly to its contour length, the denatured state is energetically more favorable. The distance between neighboring base pairs in an ss chain is greater than in the ds state due to the unstacking of base pairs. Breaking a base pair makes the molecule longer, thus reducing its potential energy, $-FL$. 

Next, we turn to the thermodynamic behavior of the system at the melting transition. In the termodynamic limit ($N\to\infty$) the free energy of the chain is dominated by the singularity of the generating function closest to the origin. Thus for the free energy per monomeric unit of the DNA chain we have 
\begin{equation}
\label{eq:free_energy}
F_N/N \sim\log z^*(T,F),  
\end{equation}
where $z^*(T,F)$ is the pole of the DNA generating function, Eq.~(\ref{eq:geomsum}).
  
\begin{figure}[htbp]
\begin{center}
\centerline{\includegraphics[width=3in]{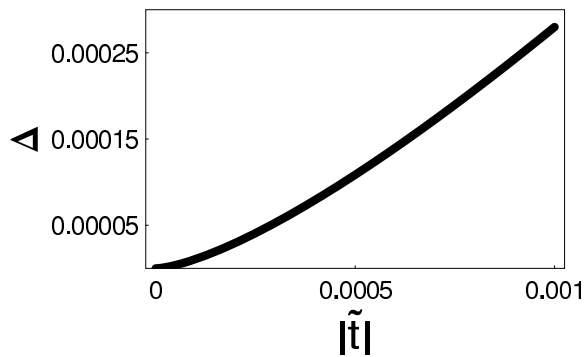}}
\caption{Numerical solution of Eq.~(\ref{eq:pole-easy}). In the vicinity of the critical point the DNA's fugacity approaches the critical fugacity as a power law $|\tilde{t}|^a$, with exponent $a\geq 1$.}
\label{fig:pole}
\end{center}
\end{figure}

To explore the behavior of the heat capacity of DNA in the close vicinity of the phase transition at $T_c(F)$  we  approximate the denominator of Eq.~(\ref{eq:geomsum}),
\begin{equation}
D(z)=1-\frac{z}{z_{ds}(T,F)}+\sigma\frac{\sqrt{1-z/z_{ss}(T,F)} }{\bigg[\log\Bigl(1-z/z_{ss}(T,F)\Bigr)  \bigg]^2} 
\end{equation}
by expanding the critical fugacity for ds segments of the chain in the first order of the reduced temperature $t=(T-T_c)/T_c$ while keeping the force $F$ as a parameter.
%We then find the pole $z^*$ by setting the denominator equal to zero
%We expand the critical fugacity of the ds and ss segments in power series of $(T-T_c)$,
%where $T_c$ is the critical temperature at which the strand separation occurs.
%Therefore, in the first order of reduced temperature $t=(T-T_c)/T_c$ we have
\begin{eqnarray}
&z_{ds}(T,F)&=z_{ds}(T_c,F)+T_c\frac{\partial z_{ds}(T,F)}{\partial T}{\bigg |}_{T_c} t
\end{eqnarray}
%At the critical temperature $T_c$ 
%\begin{equation}
%z_{ds}(T_c,F)=z_{ss}(T_c,F)\equiv z_c(F).
%\end{equation}
%Rescaling the reduced temperature, 
%\begin{equation}
%t^{old}= \frac{z_c(F)}{(\partial z_{ss}(T,F)/\partial T)\Big |_{T_c}}\, t^{new}
%\end{equation}
Thus the denominator $D(z)$ becomes 
\begin{equation}
\label{eq:denom2}
D(z)=1 - \frac{z}{z_{ds}(T_c,F)} + A(F) t+\sigma\frac{\sqrt{1-z/z_{ss}} }{\bigg[\log\Bigl(1-z/z_{ss}\Bigr)\bigg]^2}, 
\end{equation}
with
\begin{equation}
A(F)=T_c(F)\frac{\partial z_{ds}(T,F)}{\partial T}{\Big |}_{T_c(F)}  
\end{equation}
To simplify even further, we drop the term $1-z/z_c$ in Eq.~(\ref{eq:denom2}) in comparison with $A(F)t$ and rescale the reduced temperature
\begin{equation}
\tilde{t}\equiv A(F) t/\sigma,
\end{equation}
thus obtaining for the denominator $D(z)$
\begin{equation}
D(z)\propto \tilde{t} + \sqrt{\Delta}/\Big[\log\Delta\Big]^2,  
\end{equation}
where we have denoted
\begin{equation}
\Delta=1-z/z_{ss}(T,F)
\end{equation}
The pole of function $D(z)$ can be found numerically by solving the equation
\begin{equation}
\label{eq:pole-easy}
\sqrt{\Delta}/\Big[\log \Delta \Big]^2=-\tilde{t}
\end{equation}
In Fig.~(\ref{fig:pole}) we display the behavior of $\Delta$ vs reduced temperature $\tilde{t}$.
Taking the double derivative of $\Delta(t)$ in Eq.(\ref{eq:pole-easy}) with respect to reduced temperature $t$, we find for the heat capacity per monomeric unit: 
\begin{equation}
C\propto \frac{2 (\log\Delta)^5((\log\Delta)^2-24}{(\log\Delta)-4)^3}\sim (\log\Delta)^4,  
\label{eq:ceq1}
\end{equation}
with $\Delta$ being the solution to Eq.~(\ref{eq:pole-easy}).

As can be seen from the Fig.~\ref{fig:pole}, in the vincinity of the critical point $\Delta\propto|\tilde{t}|^a$ with exponent $a\geq 1$. Therefore one may conclude that the heat capacity of a DNA chain under applied stress behaves as 
\begin{eqnarray}
C\propto(\log |t| )^4.
\end{eqnarray}

%\begin{figure}[htbp]
%\begin{center}
%\centerline{\includegraphics[width=3in]{heat_cap}}
%\caption{Heat capacity per monomer (in arbitrary units) vs reduced temperature~t for various %values of the external stress. We set the cooperativity parameter $\sigma$ to 1. The DNA physical %parameters are identical to used in Fig.~\ref{fig:phase}.}
%\label{fig:heatcap}
%\end{center}
%\end{figure}
  
\begin{figure}[htbp]
\begin{center}
\centerline{\includegraphics[width=3in]{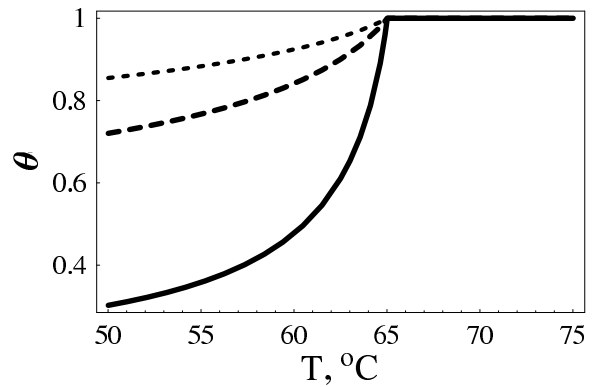}}
\caption{Fraction of denatured base pairs $\Theta$ vs temperature $T$ at the force F=65pN for various values of the cooperativity parameter $\sigma$. Dotted line corresponds to $\sigma=1$, dashed line -- to $\sigma=0.5$, dotted line -- to $\sigma=0.1$. Again, the physical parameters are the same as in Fig.~\ref{fig:phase}.}
\label{fig:fraction}
\end{center}
\end{figure}

If we denote $z_1$ and $z_2$  fugacities associated with ds and ss segments in the denominator of the DNA generating function,  
\begin{eqnarray}
&D(z_1,z_2)&=1 - z_1/z_{ds}(T,F)\nonumber\\
&&{}+\sigma\frac{\sqrt{1-z_2/z_{ss}(T,F)}}{\left[\log(1-z_2/z_{ss}(T,F))\right]^2},
\end{eqnarray}
then the fraction of denatured base pairs $\Theta$ can be determined \cite{Rudnick} as  
\begin{equation}
\frac{\Theta}{1-\Theta} = -\frac{z}{D(0,z)}\frac{\partial}{\partial z}D(0,z)
{\Bigl|}_{z^*}
\end{equation}
In Fig.~\ref{fig:fraction} we plot the fraction of denatured base pairs for various values of the cooperativity parameter $\sigma$.

\section{Informal and preliminary investigation of the results of  pulling the chains separately during thermal denaturation}
\label{sec:pulling}

In this section we briefly and incompletely consider the consequences within the Poland-Scheraga model of DNA melting of equal and opposite forces applied to the two separate chains, as illustrated in Fig. \ref{fig:stretch}b. The discussion below can be seen as a kind of  ``mean field'' extension of the PS model to take into account the concentration of the potential energy associated with stretching into the intact base pairs. A general description of the process is relatively straightforward. As denaturation bubbles form, the fraction of base pairs that remain bound diminishes, and the burden on those intact base pairs of maintaining the integrity of the duplexed form consequently increases. At some point, the forces tending to rupture the molecular bonds between those pairs exceed the threshold required to force the base pairs apart; the consequent accumulation of forces on the remaining bonded pairs induces a cascade of bond disruption, resulting in total strand separation. 

In the remainder of this section, we make use of the unmodified Poland Scheraga model to describe the statistical mechanics of melting. We append an energy term associated with pulling the chains apart, assuming that the bonds act as Hooke's law springs under the action of an external force. If one pulls on a spring with spring constant $k$ using a force with magnitude $F$, then the equation for the total energy in the spring is 
\begin{equation}
\frac{1}{2} kx^2 - Fx,
\label{eq:en1}
\end{equation}
where $x$ satisfies the equilibrium equation
\begin{equation}
F=kx.
\label{eq:en2}
\end{equation}
Inserting the solution to Eq.~(\ref{eq:en2}) into Eq.~(\ref{eq:en1}) we end up with the following expression for the total energy in the spring: $-F^2/2k$. Now, start with a set of $N$ duplexed base pairs. The original spring constant will be equal to $k_0N$. If $N_2$ of the bases have been separated, the net spring constant is reduced to $k_0(N-N_2)$. Let the force be equal to $NF_0$. This results in a total energy equal to 
\begin{equation}
-\frac{F_0^2N^2}{2(N-N_2)}.
\label{eq:en3}
\end{equation}
The partition function is now weighted by the factor $\exp \left[ F_0^2N^2/(N-N_2)\right]$, where we neglect the temperature denominator in the exponent, which can, in any case, be absorbed into a redefined $F_0$. 

\begin{figure}[htbp]
\begin{center}
\includegraphics[width=3in]{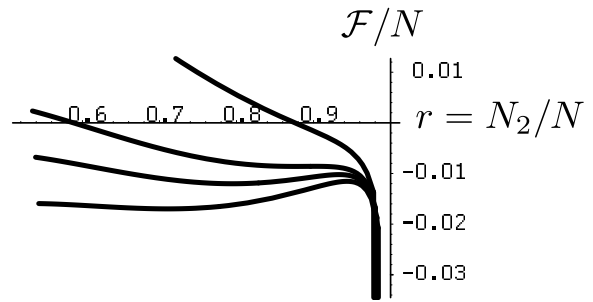}
\caption{Free energy as a function of $r$ for various values of $t$. The lower curves are for lower (more negative) $t$'s. Note a local minimum for sufficiently negative $t$ and an infinite drop in energy to the right.}
\label{fig:tearcurves}
\end{center}
\end{figure}

As for the remainder of the partition function, we obtain it from the Poland Scheraga grand partition function
\begin{equation}
\left(z_0-z_1+ct+d(z_0-z_2)^p\right)^{-1}.
\label{eq:gp1}
\end{equation}
In the above expression, $z_1$ is the fugacity of the intact pairs and $z_2$ is the same quantity for the denatured ones. The first step in calculating the partition function is to extract the coefficient in Eq.~(\ref{eq:gp1}) of $z_1^{(N-N_2)}$. We find for that coefficient
\begin{equation}
\left(z_0+ct+d(z_0-z_2)^p\right)^{-(N-N_2)},
\label{eq:gp2}
\end{equation}
where we have discarded a term of order one in the exponent. The next step is to find the coefficient of $z_2^{N_2}$ in Eq.~(\ref{eq:gp2}). Here we use the steepest descents method, looking for the extremum of 
\begin{equation}
-(N-N_2) \ln \left(z_0+ct+d(z_0-z_2)^p\right) -N_2 \ln z_2.
\label{eq:gp3}
\end{equation} 
Taking the derivative with respect to $z_2$ of Eq.~(\ref{eq:gp3}) and setting it equal to zero, we end up with the following relationship
\begin{eqnarray}
\frac{N_2}{N-N_2} &= & \frac{z_2 dp(z_0-z_2)^{p-1}}{z_0+ct +d(z_0-z_2)^p} \nonumber \\
& \equiv & X(t,z_2).
\label{eq:ext1}
\end{eqnarray}
We now introduce the variable $r=N_2/N$. Then the equation above tells us that $r=X(t,z_2)/(1+X(t,z_2))$. We can use Eq.~(\ref{eq:ext1}) as an equation giving us $z_2$ in terms of $t$ and $r$, or as an equation for $r$ in terms of $t$ and $z_2$. As our final step, we plot the total free energy, given by  
\begin{eqnarray}
\mathcal{F}&=&(N-N_2) \ln \left(z_0+ct+d(z_0-z_2)^p\right) \nonumber \\
&&{}+ N_2 \ln z_2 -\frac{N^2 F_0^2}{2(N-N_2)}
\label{eq:fe1}
\end{eqnarray}
as a function of $N_2$. We do this by first dividing (\ref{eq:fe1}) by $N$, then replacing $N_2/N$ by $r$ and, finally, plotting the result as a function of $r$. Figure \ref{fig:tearcurves} shows what this procedure leads to a range of external forces, $F_0$.  The plots are for various values of reduced temperature, $t$. Note that the curves indicate an infinite minimum of energy at $r=1$ corresponding to the lack of a limit on the lower bound of the potential energy when all bonds are broken ($N_2=N$) and the strands are pulled infinitely far apart. 

The curves in Fig. \ref{fig:tearcurves} are intended to be indicative, illustrating the presence of a metastable state of relatively intact DNA, corresponding to a minimum with respect to $r$ when the reduced temperature is sufficiently negative and a disappearance of the metastability as the reduced temperature increases. 
%%%%%%%%%%%%%%%%%%%%%%%%%%%%%%%%%%%%%%%%%%%%%%%%%%%%%%%%%%%%%%%%%%%%%%%%%%%%%%%%%%%%%%%%%%%%
\section{Conclusions} \label{sec:conclusions}

We have considered the denaturation of double stranded DNA under the applied stress within Poland and Scheraga (PS) model \cite{ps1,ps2}. In its original formulation, the PS model predicts a melting phase transition. Depending on the value of an exponent $p$, the phase transition can be first or second order. In turn, the this exponent is modified by the self-avoidance of the denaturation loop with itself and the rest of the chain \cite{kmp}.

We find that external stress transforms the action of self-avoiding interactions. In particular, because of the externally applied stress, it is energetically unfavorable for the loop to interact with the rest of the chain, while for the self-intesections within the denaturation bubble, only a subset of interaction configurations gives a sizable contribution to the loop's generating function. This results in a new analytical form for the loop's generating function, Eq.~(\ref{eq:Pi-ss}). 
As a consequence, the phase transition acquires a new signature; the heat capacity of the DNA chain behaves logarithmically in the vicinity of the phase transition (see Eq.~(\ref{eq:ceq1})). 

Another way to apply external stretching forces is to pull on opposite strands of the DNA, as shown in Fig.(\ref{fig:stretch}). In this case the applied tension energetically favors the completely denatured state at any temperature, leading either to immediate strand separation, or, under certain conditions, allowing the incompletely denatured DNA to exist as a metastable state.  If there is a transition from this metastable state to a condition of complete denaturation, it is of the spinodal type.

The key feature of biological DNA that complicates discussions of its thermal denaturation is the inherent inhomogeneity of its structure, the result of the fact that the base-pair sequence is necessarily non-uniform. As has been noted previously, and as we demonstrate in Appendix \ref{sec:harris}, the Harris criterion applies with regard to the relevance of this inhomogeneity, which can be treated as effectively random in the context of the denaturation process. Given the logarithmic modification of the specific heat at the melting transition induced by self-avoidance when there is melting under stress, we find that inhomogeneity is relevant in three dimensions and will thus alter the asymptotic behavior of the system at and in the immediate vicinity of the transition. Precisely what sort of change the inhomogeneity induces has been investigated \cite{monthus,monthus1,ares,cule}; further study will no doubt prove valuable.

\acknowledgements

We are pleased to acknowledge useful discussions with Professor Robijn Bruinsma, whose original suggestions led to the work reported here. We are also grateful to Professor Alex J. Levine for helpful interchanges. We express our gratitude to the National Science Foundation for support through grant number DMR 04-04507.

\appendix

\section{The distribution function.}
\label{sec:distribution}

The end-to-end probability distribution in Eq.~(\ref{eq:distr-F}) can be simplified to
\cite{Kleinert}  
   
\begin{eqnarray}
P_N(\boldsymbol{R},\boldsymbol{F})=\frac{i\,e^{\boldsymbol{F}\cdot\boldsymbol{R}/k_B T}}{4\,\pi^2\,l^2 R}\int_{-\infty}^{\infty}\!\!\!\! 
d\eta\, \eta\, e^{-i R\eta/l}
\left(\frac{\sin \eta}{\eta} \right)^N \!\!\!\!\! .
\label{saddle}
\end{eqnarray}

The integral in Eq.~(\ref{saddle}) can be evaluated with the use of the  saddle point approximation \cite{Kleinert}.
\begin{eqnarray}
\lefteqn{P_N(\boldsymbol{R}, \boldsymbol{F})=\frac{1}{(2\pi l^2 N)^{3/2}}
\frac{\bar{x}^2}{\rho\sqrt{1-\left(\displaystyle\frac{\bar{x}}{\sinh\bar{ x}}\right)^2}}\times}\nonumber\\
&&\displaystyle\exp\Biggl\{ N \left(\frac{F l}{k_B T}\,\rho_{\|} +\log\left[\frac{\sinh \bar{ x}}{ \bar{x}}\,e^{-\rho \bar{x}}\right]\right )\Biggr \},
\label{eq:Config2}
\end{eqnarray}
where $\rho=R/Nl$, $\rho_{\|}=R_{\|}/Nl$ is the component parallel to the force, and $\bar{x}$ is solution of the equation:

\begin{equation} 
\coth \bar{x} - \frac{1}{\bar{x}} = \rho.
\label{eq:Langevin}
\end{equation}

In the following we will use the notation  
\begin{equation}
f\equiv \frac{F l}{k_B T}.
\end{equation} 
To find where the function $P_N(\boldsymbol{R},\boldsymbol{F})$ peaks we look for the extremum of the argument
of the exponential in Eq.~(\ref{eq:Config2})
\begin{equation}
\Phi(\rho_{\|},\rho_{\perp})= f\,\rho_{\|} +\log\left[\frac{\sinh \bar{x}}{\bar{x}} e^{-\rho \bar{x}}\right]
\end{equation}
with
\begin{equation}
\rho=\sqrt{\rho_{\|}^2+\boldsymbol{\rho}_{\perp}^2}.
\end{equation}
Thus
\begin{eqnarray}
&&\frac{d\Phi(\rho_{\|},\boldsymbol{\rho}_{\perp})}{d\boldsymbol{\rho}_{\perp}}=0\ \Rightarrow\ 
\boldsymbol{\rho}_{\perp}^*=0,\\
&&\frac{d\Phi(\rho_{\|},\boldsymbol{\rho}_{\perp})}{d\rho_{\|}}=0\ \Rightarrow\   \rho^*_{\|}=\rho^{*}=\coth f -\frac{1}{f}. 
\end{eqnarray}

Expanding the function $P_N(\boldsymbol R, \boldsymbol F)$ about the maximum of $\Phi$ to the second order in $\boldsymbol{\rho}-\boldsymbol{\rho}^*$ we arrive at 

\begin{eqnarray}
&P_N(\boldsymbol{\rho},f)&\propto\left(\frac{\sinh f}{f}\right)^N\exp\left\{\frac{N}{2}\Phi^{(2)}_{\rho_{\perp},\rho_{\perp}}
\,\boldsymbol{\rho}_{\perp}^2\right\}
\times\nonumber\\
&&\exp\left\{\frac{N}{2}\Phi^{(2)}_{\rho_{\|},\rho_{\|}}\,(\rho_{\|}-\rho^*_{\|})^2\right\},  
\label{eq:distr_function1}
\end{eqnarray}
where the functions $\Phi^{(2)}_{\rho_{\|}\rho_{\|}}(f)$ and $\Phi^{(2)}_{\rho_{\perp}\rho_{\perp}}(f)$ are  
\begin{eqnarray}
\label{eq:Phi1}
&\Phi^{(2)}_{\rho_{\|}\rho_{\|}}&=\frac{d^2\Phi(\rho_{\|},\boldsymbol{\rho}_{\perp})}{d\rho_{\|}^2}
{\Biggl |}_{\rho_{\|}^*,\boldsymbol{\rho}_{\perp}^*}\!\!\!\!\!\!\!=\frac{f^2 \sinh^2\!\! f}{f^2-\sinh^2\!\! f}, \\
\label{eq:Phi2}
&
\Phi^{(2)}_{\rho_{\perp}\rho_{\perp}}&=\frac{d^2\Phi(\rho_{\|},\boldsymbol
{\rho}_{\perp})}{d\boldsymbol{\rho}_{\perp}^2}
{\Biggl |}_{\rho_{\|}^*,\boldsymbol{\rho}_{\perp}^*}\!\!\!\!\!\!\!=-\frac{f^2}{\displaystyle f\coth f-1}.
\end{eqnarray}

%%%%%%%%%%%%%%%%%%%%%%%%%%%%%%%%%%%%%%%%%%%%%%%%%%%%%%%%%%%%%%%%%%%%%%%%%
\section{Higher order corrections to the statistics of an excluded volume
 polymer under tension.}
\label{sec:higherorder}
%%%%%%%%%%%%%%%%%%%%%%%%%%%%%%%%%%%%%%%%%%%%%%%%%%%%%%%%%%%%%%%%%%%%%%%%%%
\begin{figure}[htbp]
\begin{center}
\includegraphics[width=3in]{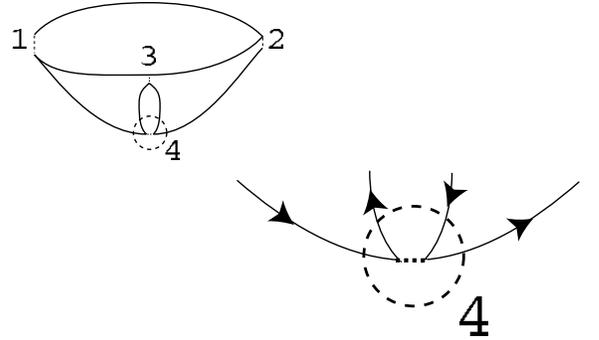}
\caption{Upper left: diagram contributing to the self-energy of the propagator for a double strand of DNA subject to an external stress. The numbers index the vertices in the diagram. Lower right: expanded view of the vertex labeled 4. The arrows indicate the ``sense'' of the strand, and the heavy dashed line is the effective repulsive interaction associated with the self-intersection occurring at the vertex. }
\label{fig:pertdiag3}
\end{center}
\end{figure}

The essence of the analysis of higher order corrections and their effect on the analytical structure of the generating function of a strand of DNA under stress is based on the argument that infrared singularities in the generating function arise from the large-distance behavior of the real space propagator and, in particular, from integrations over large separations in the multiple integrals represented by perturbation-theoretical diagrams. Consider, for instance, the diagram pictured in Fig.~\ref{fig:pertdiag3}. 
%\begin{figure}[htbp]
%\begin{center}
%\includegraphics[width=3in]{pertdiag1.eps}
%\caption{A diagram contributing to the self-energy of the propagator for a double strand of DNA subject to an external stress. The numbers index the vertices in the diagram. }
%\label{fig:pertdiag1}
%\end{center}
%\end{figure}
Consider, in particular,  the vertex of that diagram indicated by the dashed circle. The contribution to the singularity structure of that diagram of interest to us results from the integration over the position of that vertex. The influence of the external tension on the statistics of the walk is encoded in factors of the form $\exp[\boldsymbol{f}  \cdot (\boldsymbol{r}_i - \boldsymbol{r}_j)]$, where $i$ and $j$ are the indices of the vertices at the ``head`` and ``tail'' ends of a propagator line. As a given diagram consists of a single path broken into segments, each of which is a propagator line, the products of those factors is the overall $\exp[\boldsymbol{f} \cdot \boldsymbol{R}]$, where $\boldsymbol{R}$ is the displacement vector from the ``tail'' to the ``head'' of the DNA strand. This means that vertex 4 is associated with the product of factors
\begin{equation}
\exp\left[\boldsymbol{f} \cdot (\boldsymbol{r}_4 - \boldsymbol{r}_1 +\boldsymbol{r}_3-\boldsymbol{r}_4 +\boldsymbol{r}_4-\boldsymbol{r}_3 + \boldsymbol{r}_2-\boldsymbol{r}_4)\right].
\label{eq:forcefac}
\end{equation}
This tells us that the force has no direct influence on the integration over the real space position of vertex 4. The principle contributions to the integration will be of the general form
\begin{eqnarray}
\exp [ -\sqrt{6/al} \ \sqrt{1-z/z_0 +f^2a/6l}  \nonumber \\  \left(|\boldsymbol{r}_4-\boldsymbol{r}_1|+2|\boldsymbol{r}_4-\boldsymbol{r}_3|+|\boldsymbol{r}_4-\boldsymbol{r}_2|\right)],
\label{eq:pert1}
\end{eqnarray}
where multiplicative factors that are independent of $\boldsymbol{r}_4$ have been omitted. When $\boldsymbol{r}_4$ is sufficiently large, the key contributions go as $e^{  - \sqrt{6/al} \ \sqrt{1-z/z_0 +f^2a/6l}|\boldsymbol{r}_4| }$, with prefactors that are polynomial in $\boldsymbol{r}_4$. Whatever singularity results from the integration over $\boldsymbol{r}_4$ arises from the coefficient $\sqrt{1-z/z_0 +f^2a/6l}$ in the exponent, which means that the contribution to the singularity structure in the complex $z$-plane is at $z_c$ in Fig.~\ref{fig:shift}, which is, as noted previously, further from the origin than the principle singularity in the generating function for the stress-affected propagator. 

All vertices to be integrated over will be of the kind discussed immediately above, with the exception of the vertex labeled 2 in Fig.~\ref{fig:pertdiag3}, the analysis of which parallels the discussion in Section~\ref{sec:twoloop}. Thus, for this diagram---and we believe to all orders in perturbation theory---the effects of self-intersection are asymptotically negligible.

%%%%%%%%%%%%%%%%%%%%%%%%%%%%%%%%%%%%%%%%%%%%%%%%%%%%%%%%%%%%%%%%%%%%%%%%%%%%%%
\section{Demonstration of the relevance of disorder to the melting transition in DNA}
\label{sec:harris}
%%%%%%%%%%%%%%%%%%%%%%%%%%%%%%%%%%%%%%%%%%%%%%%%%%%%%%%%%%%%%%%%%%%%%%%%%%%%%%

The demonstration in this Appendix should be seen as a recapitulation and extension of the discussion by Monthus and Garel \cite{monthus}. In order to assess the effects of inhomogeneity on the statistics of the melting transition, we make use of the fact that the ``disorder'' associated with the distribution of base pairs is quenched rather than annealed, in that the base pairs are effectively frozen into place and do not rearrange in response to free energy gradients. In this case, the appropriate disorder average to take is over the free energy, or the logarithm of the partition function, rather than the partition function itself. We  then make use of the standard and useful identity 
\begin{equation}
\ln a = \lim_{n\rightarrow 0} \frac{a^{n}-1}{n}.
	\label{eq:ident}
\end{equation}

\begin{figure}[htbp]
\begin{center}
\includegraphics[width=3in]{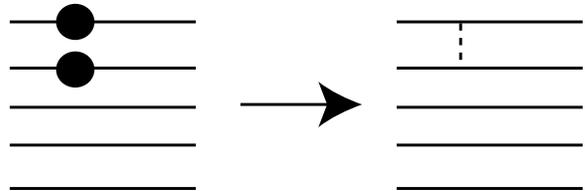}
\caption{The averaging process. The left hand side of the figure depicts the $n$ lines generated by taking the $n^{\rm th}$ power of the partition function of a melting strand of DNA, the large dots representing the disorder on a particular site. The right hand side of the figure depicts the result of the averaging of this disorder over an ensemble.  }
\label{fig:dis1}
\end{center}
\end{figure}

In order to evaluate the right hand side of (\ref{eq:ident}) we raise the partition function of denaturing DNA to the $n^{\rm th}$ power. Then, consider the disorder at a given site, which is assumed to be on average equal to zero, and which has a gaussian distribution. Figure \ref{fig:dis1} illustrates the result of the disorder averaging of this disorder.
The dots on the lines indicate the disorder that is averaged. This disorder represents a departure from the average value of, say, the binding energy of the base pairs. 

\begin{figure}[htbp]
\begin{center}
\includegraphics[width=3.0in]{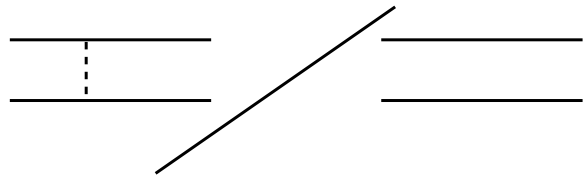}
\caption{The result of the disorder averaging of the $n^{\rm th}$ 
power of the grand partition function of denaturing DNA, after 
dividing by $n$ and setting $n=0$.}
\label{fig:dis2}
\end{center}
\end{figure}

To complete the process associated with disorder averaging, it is necessary to calculate the number of ways in which $n$ lines can be picked out and paired. This is just $n(n-1)/2$. If we divide the result of this by $n$, and then set $n$ equal to zero, we end up with the result depicted in Figure \ref{fig:dis2}.
The diagonal line in the figure represents division. The figure stands for the combination
\begin{equation}
\frac{A}{g_{0}(z)^{2}}.
	\label{eq:expression1}
\end{equation}
There are two powers of the unaveraged grand partition function in the denominator because the original expression had two powers of the partition function in the pair that is impurity averaged, and the total power of $g$ is, in the end, equal to zero.

\begin{figure}[htbp]
\begin{center}
\includegraphics[width=2in]{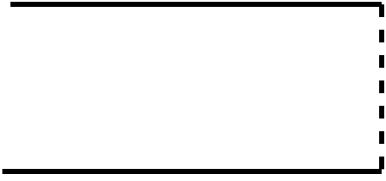}
\caption{The diagrammatic result of dividing $A$ by 
$g_{0}(z)^{2}$ as in Eq. (\ref{eq:expression1}).}
\label{fig:dis3}
\end{center}
\end{figure}

We now consider the mathematical structure of the expression $A$ in (\ref{eq:expression1}). It corresponds to two strands of DNA on each of which there is an ``impurity potential'' on one of 
the sites. Because the quantity being averages is the grand partition function, the length of the strands beyond the common site containing the impurity potential is variable. That is, for each strand in that pair, we sum over all lengths beyond the point at which there is an impurity potential. Not only that, each sum is independent. This means that we are left with two strands that have the same number of sites to the left of the impurity potential, but for which the number of sites to the right can vary. Because of this, the expression A corresponds figuratively to a digram like the one shown in Figure \ref{fig:dis3} multiplied by $g_{0}(z)^{2}$. This multiplicative factor is cancelled by the $g_{0}^{2}$ in the denominator of (\ref{eq:expression1}). Thus, the ratio in (\ref{eq:expression1}) is pictorially represented by the diagram in Fig. \ref{fig:dis3}. 

As the next step, we turn to the expression represented by Fig. \ref{fig:dis3}. The important characteristic of this expression is that the number of sites in each of the two strands leading up to the common disordered pair is equal. However, we are still performing a grand partition function sum. This means that it is necessary to work out how to extract the subset of terms in the two-strand sum corresponding to the same number of base pairs in each strand. That is, it does not suffice to simply take the product of two grand partition functions together. Progress can be made once one accepts that the principal contributions arise from the pole in the partition function that is closest to the origin.
 
We start with the Poland-Scheraga  propagator \cite{ps1,ps2}
\begin{equation}
\frac{1}{-t+(z_c-z)^p}.
\label{eq:prop1}
\end{equation}
We assume that the system is below the melting temperature; the temperature is thus explicitly taken to be negative. The pole in the propagator in (\ref{eq:prop1}) is at 
\begin{eqnarray}
z&= & z_c-t^{1/p} \nonumber \\
& \equiv & z_c(t).
\label{eq:pole1}
\end{eqnarray}
Now, let $z=z_c(t)- \Delta$. Then, the propagator becomes
\begin{eqnarray}
\frac{1}{-t+(z_c-z_c+t^{1/p} + \Delta)^p} & = & \frac{1}{-t+t(1+ \Delta t^{-1/p})^{p}} \nonumber \\
& = &  \frac{1}{p \Delta t^{1-1/p} + O(\Delta^2)} \nonumber \\
& \rightarrow & \frac{t^{(1-p)/p}}{p \Delta}.
\label{eq:res1}
\end{eqnarray}
This tells us that the residue at the pole goes as $t^{(1-p)/p}$. 

The proper combination of the two partition functions corresponding to the replicas pictured in Fig. \ref{fig:dis3} is represented as the following sum
\begin{eqnarray}
\lefteqn{\frac{t^{2(1-p)/p}}{p^2} z_c(t)^{-2} \sum_{n=0}^{\infty}\left(\frac{z}{z_c(t)} \right)^{2n}} \nonumber \\ & = & \frac{t^{2(1-p)/p}}{p^2}  \frac{1}{z_c(t)^2-z^2} \nonumber \\
& = & \frac{t^{2(1-p)/p}}{p^2}\frac{1}{(z_c(t) - z)(z_c(t) + z)}.
\label{eq:res2}
\end{eqnarray}
Given that we are interested in the behavior of this expression in the immediate vicinity of the singularity at $z=z_c(t)$we are left with a lowest-order contribution to the effect of disorder on the partition function that is proportional to
\begin{equation}
\frac{t^{2(1-p)/p}}{z_c(t)-z}.
\label{eq:dis1}
\end{equation}
If we assume that this amounts to a modification of the argument of the logarithm in the free energy, we have a new free energy that goes as
\begin{eqnarray}
\lefteqn{\ln\left(z_c(t) -z+ A t^{2(1-p)/p}\right)} \nonumber \\ & = & \ln \left(z_c-t^{1/p} +A t^{2(1-p)/p} -z \right).
\label{eq:newfree1}
\end{eqnarray}
In order that the disorder have a vanishingly small effect on the behavior of the free energy, one demands that when $t$ is very small, $t^{2(1-p)/p} \ll t^{1/p}$ or $2(1-p)/p > p$. Solving for $p$, we see that this is equivalent to requiring $p<1/2$. Now, the free energy of the model is controlled by the behavior of $z_c(t)$, which means it goes as $t^{1/p}$. Thus, the specific heat---the second temperature derivative of the free energy---goes as $t^{-2+1/p} \equiv t^{- \alpha}$. This tells us that the specific heat exponent is given by $\alpha=2-1/p$. If $p<1/2$, then $\alpha<0$. Thus, in order for the disorder to be irrelevant, we must have $\alpha<0$. Otherwise, the disorder cannot be ignored. Note that this benchmark for the relevance of disorder to the thermodynamics of the melting transition is consistent with the Harris criterion \cite{harris}.

\subsection{The effect of logarithms}

Given the results of Section \ref{sec:mutualavoidance}, we see that the power $p$ that one associates with the melting transition in the presence of a force is $p=1/2$, with a logarithmic correction. That is,  the corresponding propagator goes as 
\begin{equation}
\frac{1}{-t+(z_c-z)^{1/2}/(\ln(z_c-z))^2}.
\label{eq:prop2}
\end{equation}
The pole of this propagator will be at $z-z_c-\delta$, where
\begin{equation}
\frac{\delta^{1/2}}{(\ln \delta)^2} =t.
\label{eq:deleq1}
\end{equation}
An iterative solution to the equation (\ref{eq:deleq1}) yields 
\begin{eqnarray}
\delta & = & t^2 (\ln \delta) ^4 \nonumber \\
& \rightarrow & t^2 (\ln t^2)^4 \nonumber \\
& = & 16t^2 (\ln t)^4.
\label{eq:deleq2}
\end{eqnarray}
We then write $z=z_c-\delta - \Delta$ and calculate the residue by expanding the following in $\Delta$:
\begin{eqnarray}
\lefteqn{\left( -t+\frac{(16t^2 (\ln t)^4 + \Delta)^{1/2}}{(\ln[16t^2 (\ln t)^4 + \Delta])^2}\right)^{-1}} \nonumber \\ &\rightarrow & \left( -t+t\left(1+ \frac{1}{32} \frac{\Delta}{t^2 (\ln t)^4} \right) \right)^{-1} \nonumber \\
&\propto & \frac{t (\ln t)^4}{ \Delta}.
\label{eq:res3}
\end{eqnarray}
This tells us that the residue goes as $t (\ln t)^4$. 

Once again, we process the disorder term as in (\ref{eq:res2})--(\ref{eq:newfree1}), incorporating it into an altered argument of the logarithm, and we have a modified free energy going as
\begin{eqnarray}
\lefteqn{\ln\left(z_c(t)-z + A t^2 (\ln t)^8\right)} \nonumber \\ &=& \ln \left(z_c - t^2 (\ln t)^4 + A t^2 (\ln t)^8 -z\right).
\label{eq:newfree2}
\end{eqnarray}
It is clear that the contribution to the argument of the ``disorder'' term will dominate the shift in the pole in the ordered model when $t$ is small. This means that disorder is relevant---if just barely so---for stress-modified melting of DNA.

%\bibliographystyle{apsrev}
%\bibliography{stressbib}

\begin{thebibliography}{28}
\bibitem{psbook}
D. Poland and H.A. Scheraga,{\em Theory of helix-coil
transitions in biopolymers; statistical mechanical theory 
of order-disorder transitions in biological macromolecules},
Molecular biology (Academic Press, New York, 1970).
\bibitem{ps1}
D. Poland and H. A. Scheraga, J. Chem. Phys., {\bf 45}, 1456 (1966). 
\bibitem{ps2}
D. Poland and H. A.  Scheraga, J. Chem. Phys., {\bf 45}, 1464 (1966). 
\bibitem{pb}
M. Peyrard and A. R. Bishop, Physical Review Letters,
{\bf 62}, 23, 2755 (1989).
%5% 
\bibitem{kmp}
Y. Kafri, D. Mukamel, and L. Peliti, Physical Review Letters,
{\bf 85}, 4988 (2000).  
\bibitem {d1}
B. Duplantier, Physical Review Letters, {\bf 57}, 941 (1986).
\bibitem{d2}
B. Duplantier, Journal of Statistical Physics, {\bf 54}, 581(1989).  
\bibitem{thouless1}
D. J. Thouless,  Physical Review, {\bf 187}, 732 (1969).
\bibitem{yuvand}
P. W. Anderson and G. Yuval, Journal of Physics C (Solid State Physics),
{\bf 4}, 607, (1971). 
%10%
\bibitem{gotoh}
O. Gotoh, Y. Husimi, S. Yabuki, and A. Wada, Biopolymers,{bf 15},
655 (1976).
\bibitem{harris}
A. B. Harris, Journal of Physics C (Solid State Physics),
{\bf 7}, 1671, (1974).
\bibitem{monthus}
C. Monthus and T.Garel, European Physical Journal B, {\bf 48}, 393
(2005). 
\bibitem{bensimon1}
T. R. Strick, J. F. Allemand, D. Bensimon, and  V. Croquette,
Annual Review Of Biophysics And Biomolecular Structure, {\bf 29}, 523, (2000) 
%15 
\bibitem{lubnels1}
D. K. Lubensky, D. R. Nelson,
Physical Review Letters, {\bf 85}, 1572 (2000). 
\bibitem{lubnels2}
C. B.  Danilowicz, V. Coljee, C. Bouzig, R. S. Conroy, 
D. Lubensky, A. Sarkar, and D. R.  Nelson, M. Prentiss, Biophysical Journal, {\bf 84}, 301(A), (2003).
\bibitem{lubnels3}
C. Danilowicz, V. W. Coljee, C. Bouzigues, D. K. Lubensky, D. R. Nelson, and M. Prentiss,   Proceedings Of The National Academy Of Sciences Of The United States Of America,
{\bf 100}, 1694, (2003). 
\bibitem{RandB1}
I. Rouzina and V. A. Bloomfield,   Biophysical Journal,
{\bf 80}, 882, (2001).
\bibitem{RandB2}
I. Rouzina, V. A. Bloomfield,   Biophysical Journal,
{\bf 80}, 894, (2001). 
\bibitem{Landau:51}
L.D. Landau and E.M. Lifshitz, {\em Statistical Physics}
Vol 5. (Butterworth-Heinemann, Oxford, 1996).
\bibitem{elements}
{J. A. Rudnick, G.D. Gaspari},
{\em Elements of the random walk: an introduction for advanced students and researchers},
(Cambridge University Press, Cambridge; New York, 2004).
\bibitem{spinodal}
P. G. de Gennes, Journal of Chemical Physics, {\bf 72}, 4756 (1980).
\bibitem{Grosberg}
A. Grosberg, A. Khokhlov {\em Statistical Physics of Macromolecules},
AIP series in polymers and complex materials 
(AIP Press, New York, 1994).
\bibitem{Jeffreys} 
H. Jeffreys, {\em Methods of mathematical physics},
(University Press, Cambridge, 1972)
\bibitem{Rudnick} J. Rudnick and R. Bruinsma,
Physical Review E, {\bf 65}, 030902, (2002). 
\bibitem{monthus1}
T. Garel and C. Monthus, Journal of Statistical Mechanics --
Theory and Experiment (2005).
\bibitem{ares}
S. Ares, N.K. Voulgarakis, K.O. Rasmussen, and A.R. Bishop, 
Physical Review Letters, {\bf 94} (2005).  
\bibitem{cule}
D. Cule and T.Hwa, Physical Review Letters, {\bf 79}, 2375 (1997). 
\bibitem{Kleinert}
H.\ Kleinert {\em Path Integrals in Quantum Mechanics, Statistics, Polymer Physics, and Financial Markets} (World Scientific Pub., 2004).
%\bibitem{degennes}
%P. G. de Gennes, {\em Scaling concepts in polymer physics},
%(Cornell University Press, 1979)

\end{thebibliography}

\end{document}